\documentclass[12pt]{iopart}
\pdfoutput=1
\expandafter\let\csname equation*\endcsname\relax
\expandafter\let\csname endequation*\endcsname\relax

\usepackage{graphicx}
\usepackage{amssymb}
\usepackage{amsmath}
\usepackage{bm}
\usepackage{url}
\usepackage{epstopdf}
\usepackage{iopams}  

\begin{document}
\title{Modelling elliptically  polarised Free Electron Lasers}
\author{J R Henderson$^{1,2}$\footnote{Now at Lancaster University, Engineering, LA1 4YR, UK.}, L T Campbell$^{1,2}$, H P Freund$^{3}$ and B W J M$^{\mathrm c}$Neil$^{1}$}
\address{1 SUPA, Department of Physics, University of Strathclyde, Glasgow, G4 0NG, UK}
\address{2 ASTeC, STFC Daresbury Laboratory and Cockcroft Institute, Warrington, WA4 4AD, UK}
\address{3 Colorado State University, Fort Collins, Colorado, 80523}
\eads{\mailto{j.r.henderson@strath.ac.uk}, \mailto{lawrence.campbell@strath.ac.uk}, \mailto{b.w.j.mcneil@strath.ac.uk}}
\submitto{\NJP}

\maketitle
\newcommand{\betabold}{\boldsymbol{\beta}}
\newcommand{\vbold}{\boldsymbol{v}}
\newcommand{\xhat}{\hat{\textbf{x}}}
\newcommand{\yhat}{\hat{\textbf{y}}}
\newcommand{\zhat}{\hat{\textbf{z}}}
\newcommand{\ehat}{\hat{\textbf{e}}}
\newcommand{\fhat}{\hat{\textbf{f}}}
\newcommand{\pbold}{\textbf{p}}
\newcommand{\Ebold}{\textbf{E}}
\newcommand{\Bbold}{\textbf{B}}
\newcommand{\ubold}{\textbf{u}}
\newcommand{\xbold}{\textbf{x}}
\newcommand{\ybold}{\textbf{y}}
\newcommand{\gammabold}{\boldsymbol{\gamma}}
\newcommand{\width}{140mm}
\newcommand{\height}{110mm}

\begin{abstract}
A model of a Free Electron Laser operating with an elliptically polarised undulator is presented. The equations describing the FEL interaction, including resonant harmonic radiation fields, are averaged over an undulator period and generate a generalised Bessel function scaling factor, similar to that of planar undulator FEL theory. Comparison between simulations of the averaged model with those of an unaveraged model show very good agreement in the linear regime. Two unexpected results were found. Firstly, an  increased coupling to harmonics for elliptical rather than planar polarisarised undulators. Secondly, and thought to be unrelated to the undulator polarisation, a significantly different evolution between the averaged and unaveraged simulations of the harmonic radiation  evolution approaching FEL saturation.
\end{abstract}

\section{Introduction}
The Free-Electron Laser (FEL) is a proven source of high-power tunable radiation over a wide spectral range into the hard X-ray~\cite{McNeil xray}, where its output is transforming our ability to investigate matter and how it functions, in particular in biology~\cite{xfelsci}. In addition to the atomic spatiotemporal resolution offered by the short wavelengths and pulses, the FEL can also generate radiation output from planar through to full circular polarisation using undulators of variable ellipticity such as the APPLE-III undulator design, proposed for SwissFEL~\cite{swissund}, and the Delta undulator design~\cite{deltaund}, installed at LCLS~\cite{lclsund}. This variably polarised output offers another important degree of freedom with which to investigate the behaviour of matter and is of significant interest across a wide range of science~\cite{couprie,mazza,emma1,emma2}. FEL user facilities, such as the FERMI user facility in Italy, are now recognising and addressing this need for elliptically polarised output~\cite{fermi1,fermi2}.

In a planar undulator, the electrons have a fast axial `jitter' motion at twice the undulator period as they propagate along the undulator axis. In addition to the coupling of the electrons to the fundamental radiation wavelength, the jitter motion allows coupling to odd harmonics of the fundamental, which can also experience gain. A commonly used model used for simulating the FEL interaction is the `averaged' model which, as the name suggests, averages the governing Maxwell and Lorentz  equations describing the electron/radiation coupling over an undulator period~\cite{colson}.  The averaging of the jitter motion introduces coupling terms described by a difference of Bessel functions which depend upon both the undulator strength and the harmonic~\cite{colson,bw}. For an helical undulator, there is no electron jitter and the difference of Bessel functions coupling terms become a constant for the fundamental and zero for all harmonics, i.e in an helical undulator there is no gain coupling to harmonics. 

It is perhaps surprising that the equivalent coupling terms for an elliptically polarised undulator have not been derived previously. In this paper, the coupling terms due to electron jitter motion are calculated in a general way for all undulator ellipticities from a planar through to an helical configuration, corresponding to those now available from  variably polarised undulators, so enabling more accurate modelling of this important type of FEL output. 

The resultant derived coupling terms, which are more general form of the difference of Bessel functions factors of the planar undulator case, are used to predict the scaling of the FEL interaction for a range of undulator ellipticities. An averaged FEL simulation code then uses the general Bessel function factors to give solutions of elliptically polarised FEL output into the nonlinear, high-gain regime and tested against the scaling. A further test is also made by comparing the results of the averaged FEL simulations with an unaveraged simulation code, Puffin~\cite{Campbell Puffin}. New, perhaps unexpected, results are presented and discussed.

\section{The elliptical undulator model}
In this section the equations describing the electron beam and radiation evolution in an elliptically polarised undulator are derived in the 1D limit. The equations are averaged over an undulator period removing any sub-wavelength information or effects such as Coherent Spontaneous Emission. 

The undulator magnetic field with variable ellipticity is simply defined as:
\begin{align}
	\Bbold_u = -B_0 \sin(k_u z)\xhat + u_e B_0 \cos(k_u z)\yhat 
	\label{Ufield}
\end{align}
where $u_e$ describes the undulator ellipticity, $B_0$ the peak undulator magnetic field, and $k_u = 2 \pi/\lambda_u$ where $\lambda_u$ is the undulator period. The undulator ellipticity parameter varies in the range $0\leq u_e\leq 1$, from a planar ($u_e=0$) through to an helical undulator ($u_e=1$) to give an RMS elliptical undulator parameter of:
\begin{align}
	\bar{a}_u  = \sqrt{\frac{  1 + u_e^2}{2} }\;a_u,  \label{rms undulator parameter}
\end{align}
where the peak undulator parameter is defined as $a_u =   e B_0/m c k_u$. The resonant fundamental FEL wavelength is then: 
\begin{align}
	\lambda_r = \frac{\lambda_u}{2 \gamma_r^2}  ( 1  + \bar{a}_u^2),
\end{align}
where the resonant electron energy in units of electron rest mass $\gamma_r = \left<\gamma\right>$, the mean of the electron beam. 

\subsection{The electron equations}
In the averaged FEL model the electron orbits are first calculated in the absence of any radiation field from the Lorentz force equation:
\begin{align}
	\frac{d \betabold_j}{d t} = -\frac{e}{\gamma_j m} \betabold_j \times \boldsymbol{B}_u  
	\label{lorentz}
\end{align}
where $\betabold_j  = \textbf{v}_j/c$ and $\gamma_j$ are the $j^{th}$ electron's velocity scaled with respect to the speed of light $c$, and the corresponding Lorentz factor. Substituting for the undulator field~(\ref{Ufield}), and integrating the Lorentz equation~(\ref{lorentz}), the scaled electron velocity components are obtained:
\begin{align}
	\beta_{xj}  &=  \left({\frac{2 u_e^2 }{1 + u_e^2} }\right)^{1/2} \frac{\bar{a}_u}{ \gamma_j  }  \sin(k_u z)  \label{betax} \\ 
	\beta_{yj} &=  -\left({\frac{ 2 }{1 + u_e^2} }\right)^{1/2}  \frac{\bar{a}_u}{ \gamma_j  }  \cos(k_u z) \label{betay} \\
	\beta_{zj}  &=  \left[ \bar{\beta}_z^2  -   \left(\frac{1 -u_e^2}{1 +u_e^2}\right)\frac{\bar{a}_u^2}{  \gamma_j^2}\cos(2k_u z)\right]^{1/2} \label{betaz} \
\end{align}
where $\bar{v}_z = c\bar{\beta}_z$ is the average longitudinal electron velocity.  The  constants $m$ and $e$ take their usual meanings of rest mass and charge magnitude of the electron. Introducing the non-unit vector basis 
${\bf f} = \frac{1}{\sqrt{2} }( u_e \xhat + i \yhat)$, so that ${\bf f} \cdot {\bf f} = -(1-u_e^2)/2$ and ${\bf f} \cdot {\bf f^*} = (1+u_e^2)/2$, the perpendicular components may be written:
\begin{equation}
	{\boldsymbol \beta}_{\perp j}   =  \frac{i}{\sqrt{1+u_e^2}} \frac{\bar{a}_u}{\gamma_j}\left({\bf f}\exp\left(-ik_uz\right)- c.c.\right)  \label{betaperp}. \
\end{equation}
Integrating equation~(\ref{betaz}), the longitudinal electron trajectory in the presence of the undulator field only is:
\begin{align}
	z_j\left(t\right)  &=    c\bar{\beta}_z  t  -  \frac{ \bar{a}_u^2}{4  \gamma_j^2 k_u \bar{\beta}_z^2} \left(\frac{1 -u_e^2}{1 +u_e^2}\right)\sin(2 k_u  c\bar{\beta}_z t). \label{z oscil}
\end{align}
The oscillatory term in (\ref{z oscil}) describes the `figure-of-eight' longitudinal jitter motion of the electron in a non-helical undulator associated with coupling to harmonics of the radiation field~\cite{colson}.

A co-propagating radiation field is similarly defined using the same non-unit vector basis 
${\bf f}$ as the sum over harmonics of the fundamental resonant field, i.e.\ ${\bf E}=\sum_{n}{\bf E}_n$, where:
\begin{align}
	\Ebold_n\left(z,t\right)  = \frac{i}{\sqrt{2}}\left({\bf f}\;\mathcal{E}_n\left(z,t\right) e^{ i n(k_r z - \omega_r t)}-c.c.\right)\label{E_n}.
\end{align}
The scaled energy evolution of the $j$th electron in the transverse plane-wave radiation field of~(\ref{E_n}) may then be written as:
\begin{align}
	\frac{d \gamma_j}{d t} = - \frac{e}{mc} \sum_{n}  \betabold_{\perp j} \cdot \Ebold_n   , \label{gamma derivative wrt time1}
\end{align}
Using equations for the electron motion~(\ref{betaperp}, \ref{z oscil}), the electric field (\ref{E_n}) and the identity:
\begin{equation}
	e^{ i x \sin(\phi)} = \sum\limits_{n=-\infty}^\infty J_n(x)e^{ i n \phi},
	\label{Jidentity}
\end{equation}  
the equation for the electron energy~(\ref{gamma derivative wrt time1}) simplifies to:

\begin{align}
	\frac{d \gamma_j}{d t} = - \frac{e}{4mc} \left(\frac{2}{1+u_e^2}\right)^{1/2} \frac{\bar{a}_u}{\gamma_j} \sum_{n} \left( \mathcal{E}_n  e^{i n\theta_j  }  \left( (1+u_e^2) \right. \right. \sum\limits_{m=-\infty}^\infty J_m(b)e^{-i    (n - 1 + 2m) k_u \bar{\beta}_z c  t} \nonumber  \\ + (1-u_e^2)   \sum\limits_{m=-\infty}^\infty J_m(b)e^{-i    (n + 1 + 2m) k_u \bar{\beta}_z c  t} \left. \big)   + c.c. \right) \label{E dot beta bessel sum},
\end{align}
where $\theta_j =  ( k_r +   k_u) \bar{\beta}_z c t  -  \omega_r t$ is the pondermotive phase. Resonant, non-oscillatory terms, which do not average to zero over an undulator period occur only for $n  \pm    1 + 2m  = 0$, so that on averaging over an undulator period equation~(\ref{E dot beta bessel sum}) simplifies further to:
\begin{align}
	\frac{d \gamma_j}{d t} = - \frac{e}{4mc} \left(\frac{2}{1+u_e^2}\right)^{1/2} \frac{\bar{a}_u}{\gamma_j} \sum_{n} JJ_n  \big(   \mathcal{E}_n  e^{i n\theta_j  }    \  + c.c. \big),  \label{E dot beta bessel} \
\end{align}
where: 
\begin{align}
	JJ_n &= (-1)^{\frac{n-1}{2}} \big( (1+u_e^2) J_{\frac{n-1}{2}}(\xi) - (1-u_e^2)  J_{\frac{n+1}{2}}(\xi) \big), \\
	\xi &=    \frac{n  \bar{a}_u^2}{2(1 + \bar{a}_u^2)  } \frac{ 1 -u_e^2}{ 1 + u_e^2}.
\end{align}


\subsection{The wave equation}
The 1D wave equation is used to model the plane wave radiation field evolution and is given by:
\begin{align}
	\Big( \frac{\partial^2 }{\partial z^2} - \frac{1}{c^2} \frac{\partial^2 }{\partial t^2} \Big) \Ebold = \frac{\mu_0}{\sigma} \frac{\partial \textbf{J}_\perp}{\partial t} \label{Maxwell wave equation}
\end{align}
where $\sigma$ is the transverse area of the co-propagating planar radiation field and electron beam with transverse current density of ${\bf J}_\perp=-ec\sum_{j=1}^{N}\betabold_{\perp}\delta\left({\bf r}-{\bf r}_j\left(t\right)\right)$.
The transverse components of the electric field and transverse current density are defined by $E_{\perp} = \sqrt{2}\;  \Ebold\cdot {\bf f^*}$ and $J_{\perp} = \sqrt{2}\; \textbf{J} \cdot{\bf f^*} $ respectively.  In the 1D limit, the wave equation~(\ref{Maxwell wave equation}) simplifies to:
\begin{align}
	\Big( \frac{\partial^2 }{\partial z^2} - \frac{1}{c^2} \frac{\partial^2 }{\partial t^2} \Big)  E_{\perp}  = \frac{\mu_0}{\sigma} \frac{\partial  J_{\perp}}{\partial t}.  \label{1D Maxwell Wave Equation}
\end{align}
By using the transverse velocity~(\ref{betaperp}), the harmonic fields~(\ref{E_n}) and by neglecting the backward wave as detailed in~\cite{bw} then, using the Bessel identity~(\ref{Jidentity}), the wave equation~(\ref{1D Maxwell Wave Equation}) reduces to a wave equation for each harmonic envelope $\mathcal{E}_n$:
\begin{align}
	\Big( \frac{\partial   }{\partial z } + \frac{1}{c} \frac{\partial}{\partial t} \Big) \mathcal{E}_n  
	=  
	\frac{e\mu_0 c^2\bar{a}_u}{\sqrt{2}\sigma   (1+u_e^2)^{3/2}}     
	\sum_{j=1}^N     
	\frac{e^{-in \theta_j}}{\gamma_j}   \big( (1+u_e^2)    \sum\limits_{m=-\infty}^\infty J_m(b) e^{i  (n-1 + 2m) k_u \bar{\beta}_z c    t}   \nonumber \\ + (1- u_e^2)   \sum\limits_{m=-\infty}^\infty   J_m(b)e^{i (n+1 + 2m)k_u   \bar{\beta}_z c   t  }      \big)    \ \delta(z - z_j(t)),   \label{Wave Equation sum of Bessel function}
\end{align}
where $\theta = (k_r + k_u)z - \omega_r t$ is the ponderomotive phase of the fundamental wavelength. Resonant terms are only seen to occur for $ n \pm 1 + 2m  = 0$ and, as $m$ is integer,  the harmonic numbers $n$ are therefore odd. Applying this resonant condition yields:
\begin{align}
	\Big( \frac{\partial   }{\partial z } + \frac{1}{c} \frac{\partial}{\partial t} \Big) \mathcal{E}_n  
	=  \frac{e\mu_0 c^2\bar{a}_u}{\sqrt{2}\sigma   (1+u_e^2)^{3/2}}    \sum_{j=1}^N    \frac{e^{-in \theta }}{  \gamma_j}     JJ_n    \ \delta(z - z_j(t)).
	\label{uawave}
\end{align}

\subsection{The scaled FEL model}
The scaling of~\cite{bnp,sr} is now applied using the FEL parameter $\rho =  \gamma_r^{-1} (\bar{a}_u \omega_p / 4 c k_u )^{2/3}$, where $\omega_p$ is the peak  non-relativistic plasma frequency of the electron beam. The wave equation for field~(\ref{uawave}) is also averaged over a radiation wavelength by assuming the field envelope does not change in this interval.
The independent variables are the scaled distance through the FEL $\bar{z} = z/l_g$,  and scaled position in the electron beam rest-frame $\bar{z}_1 =  (z - c \bar{\beta}_z t)/ \bar{\beta}_zl_c = 2 \rho \theta_j$, where $l_g=\lambda_u/4\pi\rho$ and $l_c=\lambda_r/4\pi\rho$ are respectively the gain length and cooperation length of the FEL interaction at the fundamental  ($n=1$) in an helical undulator ($u_e=1$)~\cite{sr}. Clearly, and as shown from the scaling below, these lengths are different for interactions at harmonics and in an elliptical undulator.

Introducing the scaled harmonic radiation envelopes:
\begin{align}
	A_n = \sqrt{\frac{1+u_e^2}{2}} \frac{\bar{a}_u e \mathcal{E}_n}{4 m c^2 k_u \left( \rho \gamma_r \right)^{2}},
	\label{A_n}
\end{align}
the scaled electron energy $p_j = (\gamma_r - \gamma_j)/\rho \gamma_r$ and using the definition of the ponderomotive phase $\theta$, the scaled equations for the 1D FEL interaction in an elliptically polarised undulator including harmonic radiation fields are given by:
\begin{align}
	\frac{d \theta_j}{d \bar{z}} &= p_j \label{dtheta} \\
	\frac{d  p_j}{d\bar{z}}  &=  -\sum_{n, odd} \alpha_n \Big( A_n  e^{i \theta_j}  + c.c. \Big) \label{dp}\\
	\left( \frac{\partial}{\partial \bar{z}} + \frac{\partial}{\partial \bar{z}_1} \right)  A_n   &=   \alpha_n\; \chi\left(\bar{z}_1\right)\big<e^{- i \theta_j} \big>.\label{dA} \
\end{align}
where $\alpha_n$ are ellipticity dependent coupling parameters given by:
\begin{align}
	\alpha_n = \frac{JJ_n}{1 + u_e^2},
	\label{alpha}
\end{align}  
and $\chi\left(\bar{z}_1\right)=I\left(\bar{z}_1\right)/I_{pk}$ is the beam current scaled with respect to its peak value~\cite{bw}. There is one wave equation of type~(\ref{dA}) for each harmonic considered.
Notice from~(\ref{A_n}), that the harmonic field envelopes $\mathcal{E}_n$ are scaled so that the $|A_n|^2$ are proportional to the power of the elliptically polarised harmonic radiation fields over the full range of $u_e$, from planar to helical polarisation.

\section{Modelling the elliptical undulator FEL}

The equations for the elliptical model~(\ref{dtheta} - \ref{dA}) are now solved for a range of ellipticity parameters $u_e$. 
The solutions are determined by the ellipticity and harmonic dependent coupling parameters $\alpha_n$ which are specified and used in scaling to predict the gain length and saturation powers of the elliptical FEL interaction. 

Numerical solutions of the averaged elliptical FEL model of above are also compared with the unaveraged model of `Puffin'~\cite{Campbell Puffin}. As the  equations of this model are unaveraged, no factors such as~(\ref{alpha}) appear in the model and Puffin can simulate the FEL interaction for an undulator of any ellipticity and over a broad radiation bandwidth that includes harmonic content.
\subsection{Scaling}
\begin{figure}[!ht]
	\centering
	\includegraphics[width=\width,height=120mm]{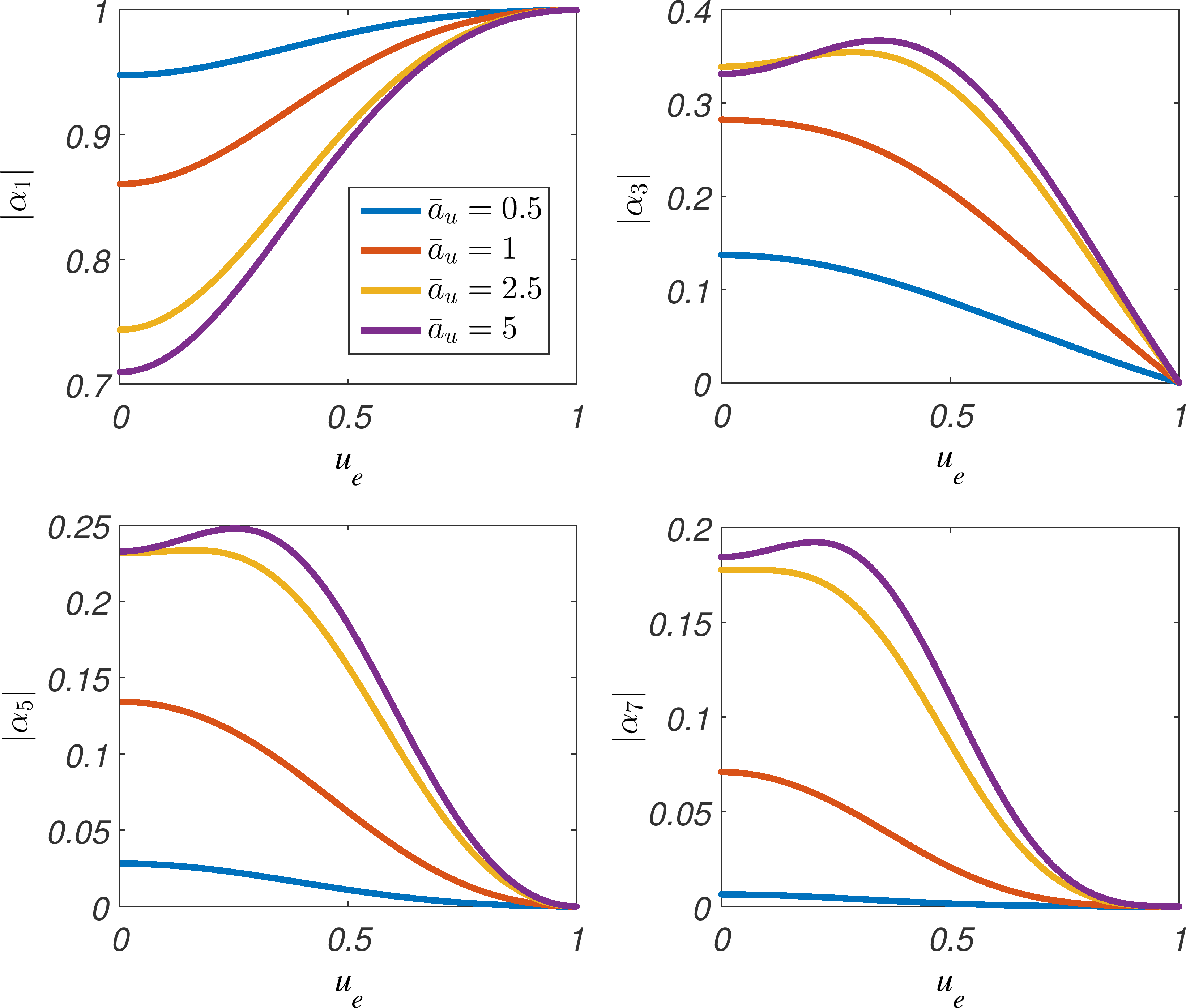}
	\caption[]{The elliptical coupling parameters $\alpha_n$ plotted as a function of the ellipticity parameter $u_e$ for the first four odd harmonics $n=1, 3 ,5 ,7$. Four different RMS undulator parameters are shown in each graph: $\bar{a}_u=0.5 , 1.0, 2.5, 5.0$. The $\alpha_n$ agree with previous analysis in the helical and planar limits, $u_e=1$ and $u_e=0$ respectively.
		Note that for larger undulator parameters $\bar{a}_u$, the coupling parameters $\alpha_n$ for harmonics maximise for an elliptical undulator configuration, $u_e>0$. For example, for the third harmonic with $\bar{a}_u=5$, then $\alpha_3$ is maximised for an undulator ellipticity of $u_e \approx 0.34$. 
\label{fig1}} 
\end{figure}

Figure~(\ref{fig1}) plots the elliptical coupling parameters $\alpha_n$ as a function of the ellipticity parameter $u_e$ for the resonant odd harmonics $n=1..7$ and for a range of RMS undulator parameters $\bar{a}_u$. 
The coupling parameters agree with previous results in the helical and planar limits. It is worth noting that for the harmonic fields $n>1$, and for larger undulator parameters $\bar{a}_u$, that the coupling is stronger for elliptically polarised undulators rather than the planar case of $u_e=0$. This result is perhaps somewhat unexpected.

If the equations for the elliptical model~(\ref{dtheta}-\ref{dA}) are written in the absence of any harmonic interactions, i.e. for $n=1$ only, then the elliptical coupling parameter $\alpha_1$ could be incorporated into the scaling to give a system of universally scaled equations with no free parameters~\cite{bnp}. In this case the FEL scaling parameter would now depend upon the elliptical coupling parameter for the fundamental as $\rho \propto  \alpha_1^{2/3}$, so that  the gain length of the interaction, and so also the saturation length $z_{sat}$, would scale as $l_g, z_{sat} \propto \alpha_1^{-2/3}$. The scaled saturation power would scale as $|A|^2_{sat} \propto \alpha_1^{2/3}$. 

In the  simulations which follow, an electron pulse of charge 70 pC is assumed with a uniform current, $\chi(\bar{z}_1)=1$, over scaled pulse length of $\bar{l}_e = l_e/l_c  = 129$. A mean beam energy $\gamma_r = 1500$ with  zero energy spread and an FEL parameter of $\rho = 2 \times 10^{-3}$ is used. Unless otherwise stated, the undulator has fixed RMS undulator parameter of $\bar{a}_u = 1.0$ independent of the undulator ellipticity, to give a fixed resonant radiation wavelength of $\lambda_r=16$nm. A seed laser of scaled amplitude of $A_0 = 10^{-4}$ was used to initiate the FEL interaction. This eliminates shot-to-shot variation of the radiation pulse saturation energy and saturation length which occurs when the interaction starts from noise, simplifying comparison with analysis and the results obtained from the solutions of the different numerical codes. The total scaled energy of an harmonic of the radiation pulse is defined by:
\begin{align}
	E_n(\bar{z})=\int^{+\infty}_{-\infty} |A_n(\bar{z},\bar{z}_1)|^2 d \bar{z}_1.
	\label{energy}
\end{align}
with the total given by the sum over the odd harmonics $E=\sum_{n,odd}E_n$.
As the electron pulse is many cooperation lengths long ($\bar{l}_e = 129$) and the interaction is seeded, the interaction will approximate a steady-state interaction where pulse effects are small. In this case, the scaled pulse energy at saturation, either for a particular harmonic component $n$ or for the total,  will be $E_{sat}\approx \bar{l_e} |A|_{sat}^2$. For an helical undulator in the steady-state, the scaled saturation power of the fundamental ($n=1$) is $|A|_{sat}^2\approx 1.37$. For the case considered here this gives a scaled pulse energy at saturation of $E_{sat}\approx 177$. 

\begin{figure}
	\includegraphics[width=\width,height=\height]{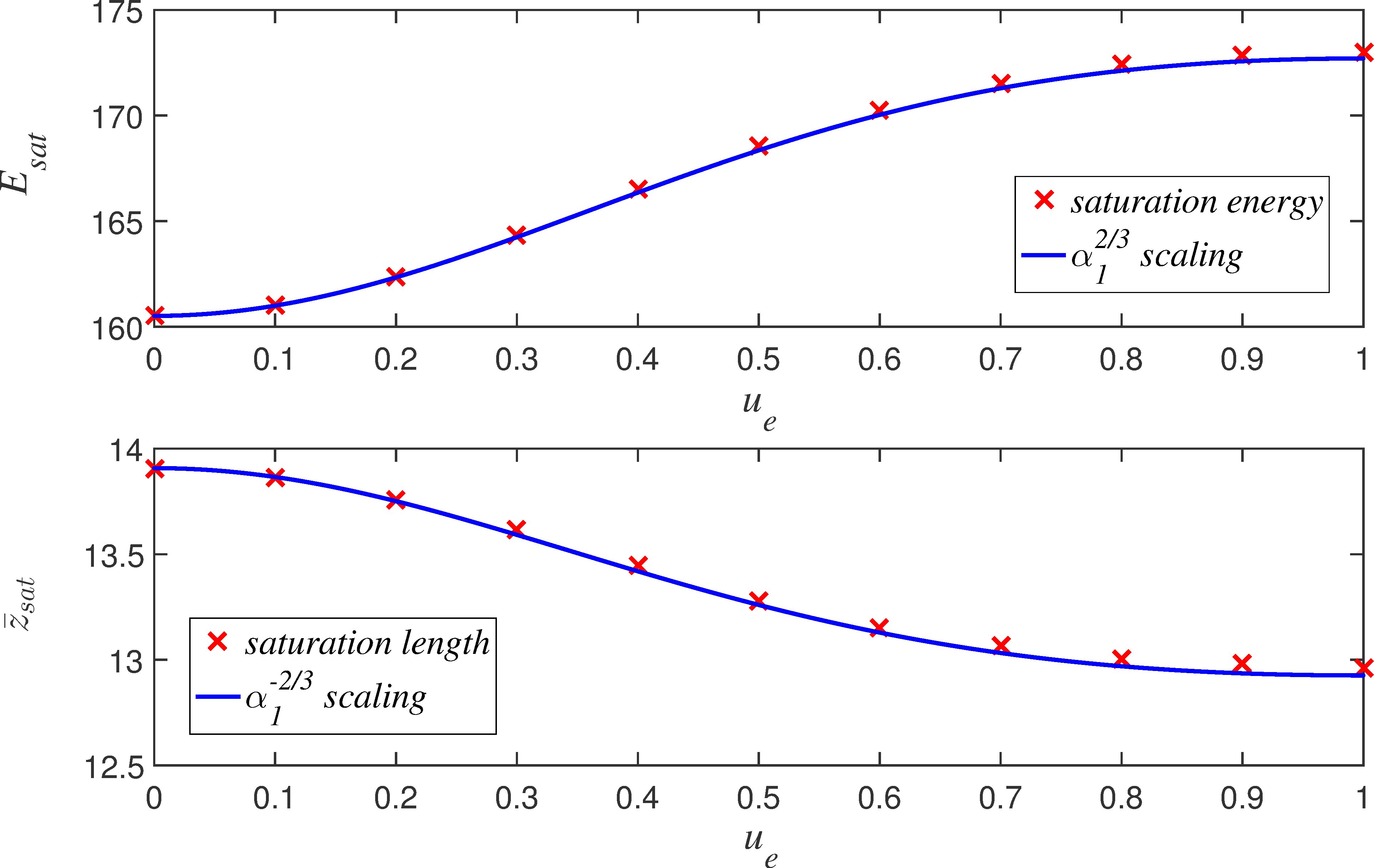}
	\caption[]{Comparison between numerical solutions of the averaged model of equations~(\ref{dtheta} - \ref{dA}) in the absence of any harmonic interactions (red crosses) and the predicted scaling with respect to the elliptical coupling parameter of the fundamental $\alpha_1$ (blue line) for the full range of the ellipticity from planar ($u_e=0$) to helical ($u_e=1$). The top plot shows the saturated pulse energy $E_{sat}$ and the lower the scaled saturation length $\bar{z}_{sat}$.
\label{fig2}} 
\end{figure}

In order to test the above scaling for the scaled saturation energy and saturation length, the equations~(\ref{dtheta} - \ref{dA}) were solved numerically for the above parameters in the absence of any harmonic interaction for a range of undulator ellipticities.   Figure~\ref{fig2} demonstrates that the numerical solutions are in very good agreement with the predicted scaling.

\subsection{Comparison between averaged and unaveraged models}
Numerical solutions to the  averaged elliptical model of equations~(\ref{dtheta}-\ref{dA}) are now compared with the those generated by the unaveraged code Puffin~\cite{Campbell Puffin}, which is able to model an FEL interaction in an elliptically polarised undulator across a broad bandwidth radiation field that includes harmonic content. The unaveraged electron motion of the Puffin model includes any `jitter' motion of equation~(\ref{z oscil}) due to an elliptically polarised undulator.

As Puffin is an unaveraged FEL simulator, the effects of Self Amplified Coherent Spontaneous Emission (SACSE) can be significant when modelling a `flat-top' electron bunch which has discontinuities in the electron beam current. As these effects cannot be modelled in an averaged model, the electron bunch used in the Puffin simulations here is modified to have smooth ramp down in current over several radiation wavelengths at the electron bunch edges. This smooth ramping of the current significantly reduces the generation of any Coherent Spontaneous Emission, enabling a better comparison between the two models.

In what follows, only the fundamental and third harmonics ($n=1,3$) are modelled using the above parameters. In the averaged model, the harmonic radiation content is obtained directly from the individual harmonic components, $A_n$. In the unaveraged model, however, access to the content of each harmonic is obtained by fourier filtering the  broadband radiation field about a narrow bandwidth of the particular harmonic of interest (in this case for $n=3$.)

\begin{figure}[!ht]
	\centering
	\includegraphics[width=\width,height=\height]{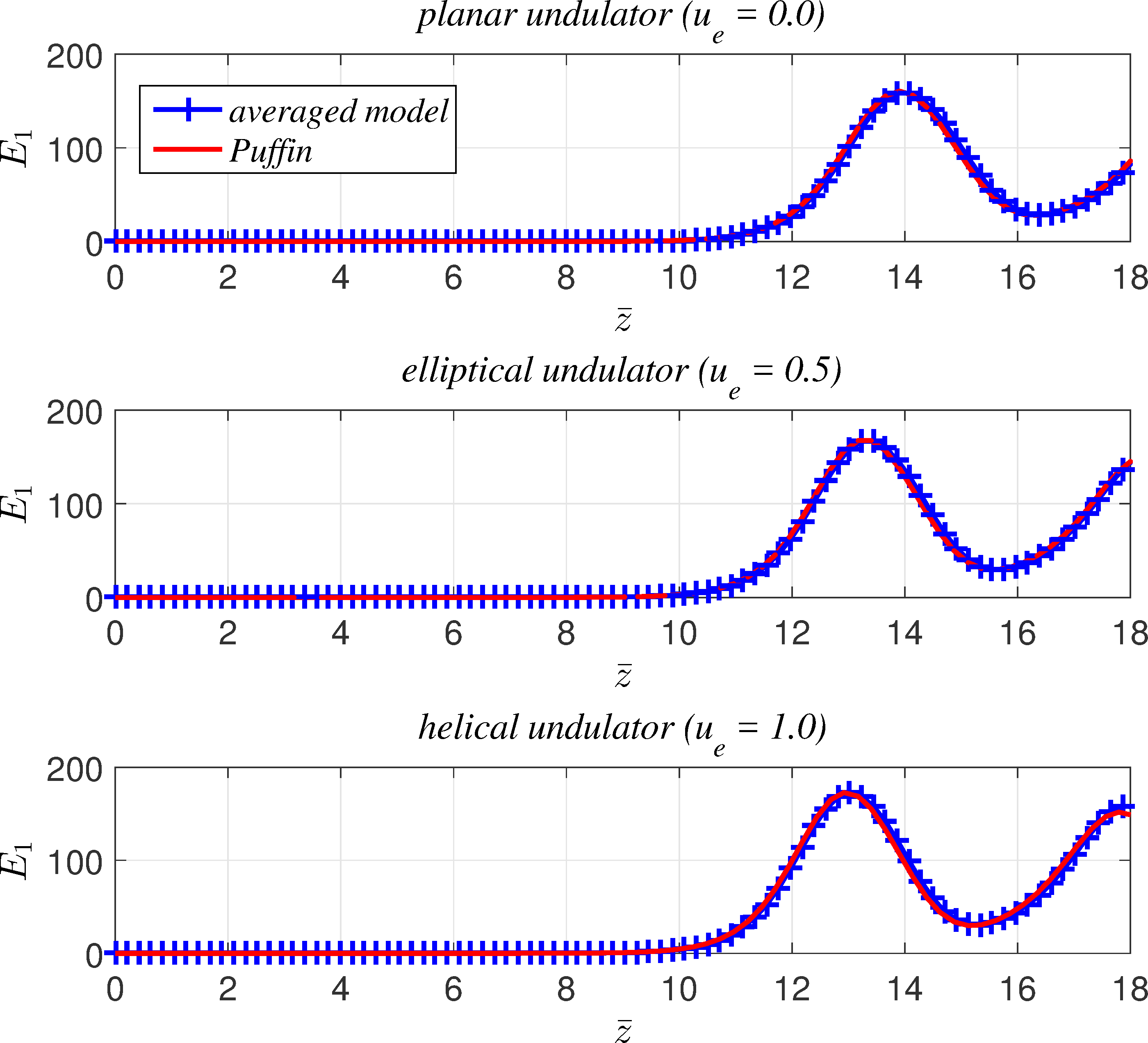}
	\caption[]{Simulations using the averaged and unaveraged models show excellent agreement for the evolution of the scaled radiation pulse energy of the fundamental $E_1$, as a function of scaled distance through the undulator for planar (top, $u_e=0.0$), elliptical (middle, $u_e=0.5$) and helical (bottom, $u_e=1.0$) undulator polarisation. \label{fig3}} 
\end{figure}

Figure~\ref{fig3} plots the scaled pulse energy of the fundamental $E_1$, from the averaged and Puffin simulations as a function of scaled propagation distance through the interaction $\bar{z}$, for three different undulator ellipticities, $u_e = 0, 0.5, 1.0$.  
Excellent agreement  between the simulations is seen for all $u_e$, well into the saturated, non-linear regime. 

\begin{figure}[!ht]
	\centering
	\includegraphics[width=\width,height=\height]{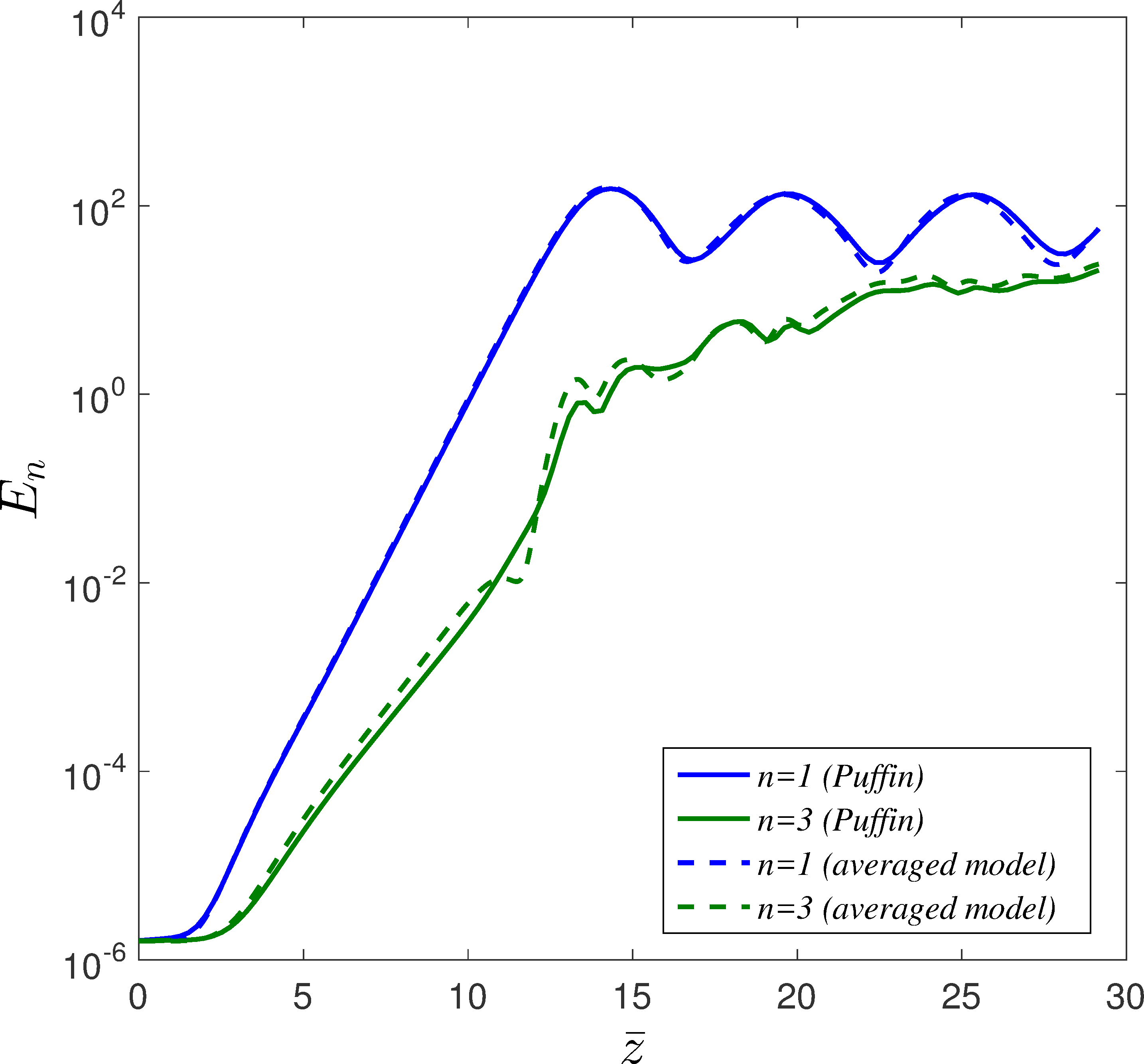}
	\caption[]{Comparison of the scaled pulse radiation energies $E_{1,3}$ for averaged and unaveraged simulations in a planar undulator ($u_e=0$), of RMS undulator parameter $\bar{a}=1.0$. Good agreement is seen except in the interval $11<\bar{z}<14$.
		\label{fig4}} 
\end{figure}

The scaled radiation pulse energies $E_n$ of the fundamental and third harmonic for both averaged and unaveraged simulations for the planar undulator ($u_e=0$) are shown in  figure~\ref{fig4}. 
As previously seen in figure~\ref{fig3}, the fundamental pulse energies $E_1$ of the averaged and unaveraged simulations are in excellent agreement.
The third harmonic shows reasonable agreement in the decoupled linear regime until $\bar{z} \approx 11$. At this point in the averaged model, the electron bunching at the fundamental also begins to drive the third harmonic field with a growth rate $\sim 3$ times that of the fundamental~\cite{harb}. While there is evidence of similar enhanced harmonic growth in the unaveraged simulation, the effect is seen to be significantly less pronounced. As the interaction proceeds into the non-linear, saturation regime for $\bar{z}>13$, both simulations are seen to resume a similar evolution. 

It was noted from figure~\ref{fig1} that for larger undulator parameters $\bar{a}_u$, the coupling parameters $\alpha_n$ for harmonics maximise for an elliptical undulator configuration, $u_e>0$. This increased coupling can be expected to decrease the gain length and increase the saturation pulse energies of harmonics for these elliptical polarisations. In particular, the gain length for the third harmonic in an undulator with parameter $\bar{a}_u = 5.0$, should be minimised for an elliptical undulator with $u_e \approx 0.34$. From the above scaling (and writing $l_g(u_e)$, etc) the ratio of the two gain lengths $l_g(0.34)/l_g(0)= 0.934$. 

\begin{figure}[!ht]
	\centering
	\includegraphics[width=\width,height=\height]{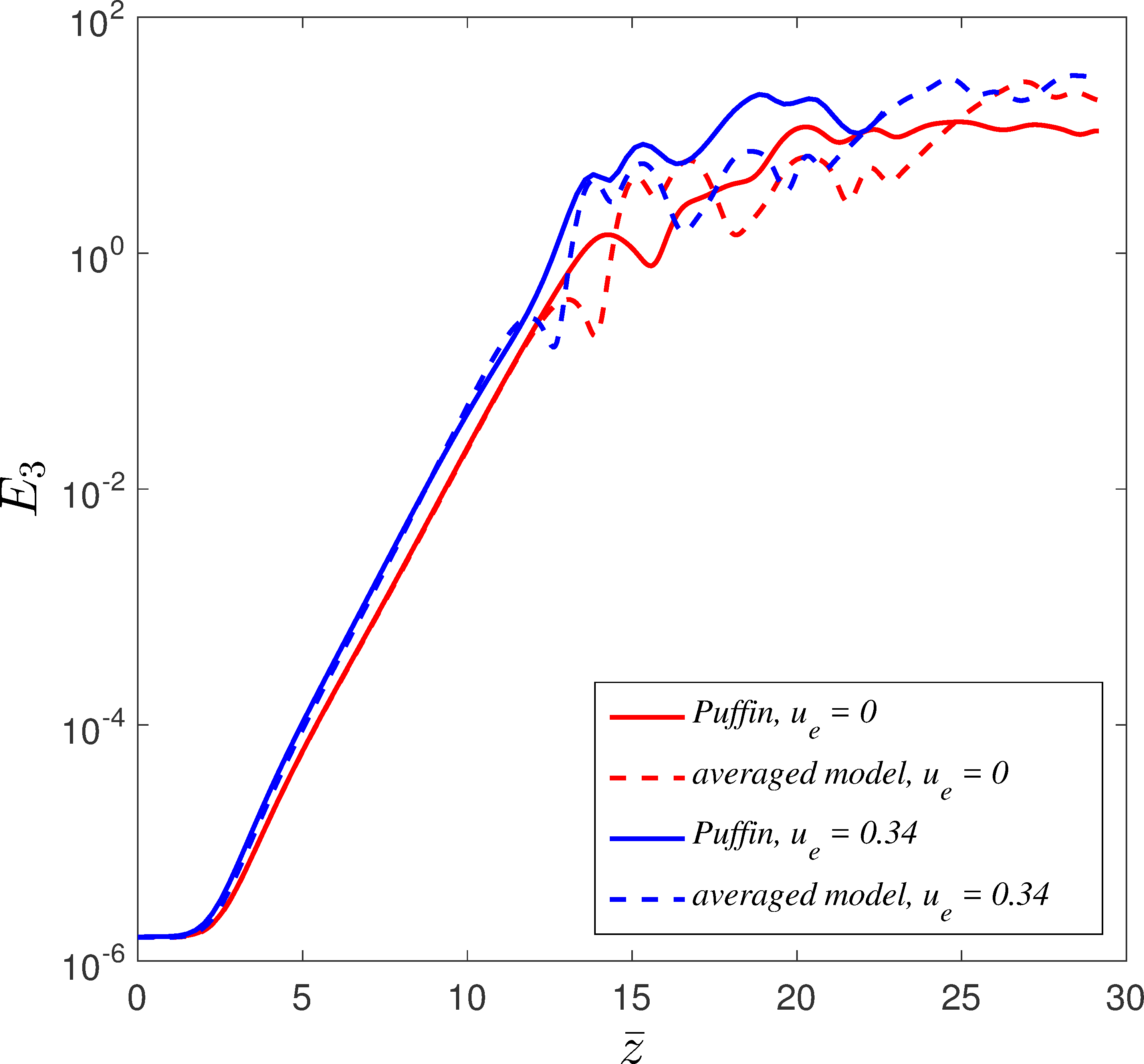}
	\caption[]{ Comparison of the scaled pulse energies for both averaged and unaveraged simulations of the third harmonics $E_3$ in an undulator with $a_w = 5.0$ for two different undulator ellipticities $u_e=0.0$ (planar undulator) and $u_e = 0.34$ (elliptical undulator). The third harmonic interaction is see to be stronger for the elliptical undulator, in agreement with the results of figure~\ref{fig1}, which shows that the  coupling parameter is maximum for the elliptical undulator case. The gain lengths of both results agree well with predicted scaling via the elliptical coupling parameter $\alpha_3$. 
		\label{fig5} } 
\end{figure}

Both the averaged and unaveraged numerical models were also used to simulate both undulator ellipticities $u_e=(0, 0.34)$ for the same value of $\bar{a}_u = 5.0$. The results are shown in figure~\ref{fig5}. The simulations are seen to agree well with each other in the linear regime with the elliptical undulator measured as having the shorter gain length $l_g(0.34)/l_g(0)\approx 0.931$, in good agreement with the value calculated from  scaling.

A similar scaling argument for the electron pulse energies at saturation gives $E_3(0.34)/E_3(0)=1.071$ which is more difficult to compare with the simulations of figure~\ref{fig5} due to the problem in defining the points of saturation. 
Note again, the difference in the simulation results between the averaged and unaveraged models as saturation is approached and the fundamental interaction drives that of the harmonic. The divergence between the two models is probably more pronounced in this case where $\bar{a}_u = 5.0$, than that of figure~\ref{fig4} where $\bar{a}_u = 1.0$.

\newpage
\section{Conclusion}
An averaged FEL model in the 1D limit for ellipticity polarised undulators including resonant radiation harmonics was presented. The  undulator ellipticity changes the previous difference of Bessel functions factor, familiar from planar undulator FEL theory, into a more general elliptical Bessel function factor, valid for a planar undulator through to an helical undulator. This new elliptical factor was incorporated into a set of averaged, scaled, differential equations describing the FEL interaction. The scaling of these equations  allows important quantities such as the gain length and radiation pulse energy, to be estimated as a function of the undulator ellipticity. 

This averaged elliptical FEL model of the  undulator  was also solved numerically and the scaling demonstrated. One notable result is that the harmonic gain and saturation energy for larger values of the undulator parameter $\bar{a}_u$, was greater for elliptically polarised undulators than for the planar equivalent.

The averaged elliptical FEL model was also compared with the numerical simulations of an unaveraged FEL model using the Puffin code which is also able to model elliptically polarised undulators (also in 3D). Overall, there was very good agreement between the two models. However, there were differences noted in the radiation pulse energy evolution of the harmonics as the interactions approached saturation and the harmonics are strongly coupled and  driven by the interaction at the fundamental. This is not directly related to the ellipticity  of the polarisation, but is thought to be a more general issue related to the validity of the averaging process in accurately describing the coupling between the fundamental and harmonic interactions. This topic will require further research. 

\ack
We gratefully acknowledge support of Science and Technology Facilities Council Agreement Number 4163192 Release \#3; 
ARCHIE-WeSt HPC, EPSRC grant EP/K000586/1;
EPSRC Grant EP/M011607/1; and
John von Neumann Institute for Computing (NIC) on JUROPA at Jlich Supercomputing Centre (JSC), under project HHH20

\end{document}